\documentclass{article}
\usepackage{arxiv}

\usepackage[utf8]{inputenc} 
\usepackage[T1]{fontenc}    
\usepackage{hyperref}       
\usepackage{url}            
\usepackage{booktabs}       
\usepackage{lipsum}		

\usepackage{doi}

\usepackage{graphicx}%
\usepackage{multirow}%
\usepackage{amsmath,amssymb,amsfonts}%
\usepackage{amsthm}%
\usepackage{mathrsfs}%
\usepackage[title]{appendix}%
\usepackage{xcolor}%
\usepackage{textcomp}%
\usepackage{booktabs}%
\usepackage{algorithm}%
\usepackage{listings}
\title{A template for the \emph{arxiv} style}

\begin{document}
\vspace*{0.35in}

\begin{flushleft}
{\Large
\textbf\newline{Optical and Electrical Properties  of Diamond-like-Carbon Coatings Prepared by Electron Cyclotron Resonance Ion Beam Deposition Process.}
}
\newline
\\
Callum Wiseman\textsuperscript{1},
Marwa Ben Yaala\textsuperscript{1,*},
Chalisa Gier\textsuperscript{1},
Laurent Marot\textsuperscript{2},
Christopher McCormick\textsuperscript{1},
Caspar Clark\textsuperscript{3},
Sheila Rowan\textsuperscript{4},
StuartReid\textsuperscript{1}
\\
\bigskip
\bf{1} SUPA, Department of Biomedical Engineering, University of Strathclyde, 106 Rottenrow East, G4 0NW, Glasgow, United Kingdom
\\
\bf{2} Department of Physics, University of Basel, Klingelbergstrasse 82, 4056, Basel, Switzerland
\\
\bf{3} SUPA, Institute for Gravitational Research, University of Glasgow, University Avenue, G12 8QQ, Glasgow, United Kingdom
\\
\bf{4} Helia Photonics Limited, Rosebank Technology Park, EH54 7EJ, Livingston, West Lothian, United Kingdom
\\
\bigskip
* marwa.ben-yaala@strath.ac.uk

\end{flushleft}

\begin{abstract}
Diamond-like carbon thin films have emerged as durable, chemically stable optical coatings for many optical and optoelectronics applications due to their hardness, chemical inertness, and optical transparency. This paper presents a novel high-energy electron cyclotron resonance ion beam sputter deposition technique to fabricate pure diamond-like carbon coatings  at room temperature. 
The chemical composition of the deposited coatings including ratios of $sp^2$/$sp^3$ bonding in the thin films were determined by X-ray photoelectron spectroscopy. Results indicate that the $sp^3$ percentage ranges from 45\% - 85\%. The transmission and reflectance spectra of the coatings were measured from UV to IR ($\lambda$ = 185 to 2500 nm) by utilizing a spectrophotometer. 
The measured spectra were analysed by the Tauc method to determine the optical band gap and Urbach energy and an optical fitting software, which utilizes the model modified by OJL, to extract the refractive index and extinction coefficient. 
By varying the ion energy, the optical properties were found to be n = 2.30 - 2.51 at 550~nm, band gap energy = 0.4 - 0.68~eV, and the Urbach energy = 0.33 - 0.49~eV. 
This study provides a flexible method for tuning the structural, optical, and electronic properties of diamond-like carbon coatings by controlling the ion energy during deposition.
\end{abstract}


\section{Introduction}\label{sec1}

Diamond-Like Carbon (DLC) coating is a class of amorphous carbon material containing both trigonally and tetrahedrally (\textit{$sp^{2}$}  and \textit{$sp^{3}$} respectively) bonded carbon atoms. DLC coatings are desirable for their varied and beneficial properties that can be controlled by changing the \textit{$sp^{2}$}/\textit{$sp^{3}$ } ratio as well as their ability to be tuned to meet the unique needs of various applications. DLCs can exhibit many properties, including high mechanical hardness, chemical inertness, high electrical resistivity, and optical transparency in the IR, as well as low friction, corrosion and wear resistance. These unique properties make DLCs suitable as a protective, chemically stable coating for many optical and opto-electronics applications \cite{Yang2021, Semikina2001, Choi2008, Szmidt1994, Litovchenko1997}. DLC films are produced by several techniques including filtered cathodic vacuum arc deposition \cite{Wu2007} ; RF magnetron sputtering \cite{Sanchez2000} ; RF plasma enhanced chemical vapour deposition \cite{Cuong2003} and RF ion beam deposition \cite{Palshin1995}. This paper presents a room temperature, high energy sputtering technique for producing DLC coatings with tuneable \textit{$sp^{3}$} content along with analysis of various properties including structural properties; optical transmittance and reflectance; refractive index and extinction coefficient; optical band gap and Urbach energy as well as electrical properties of the produced coatings.

\section{Experimental} \label{experimental}
\subsection{ECR Ion Beam Deposition process}
\begin{figure}[H]
\centering
\includegraphics[width=.7\textwidth]{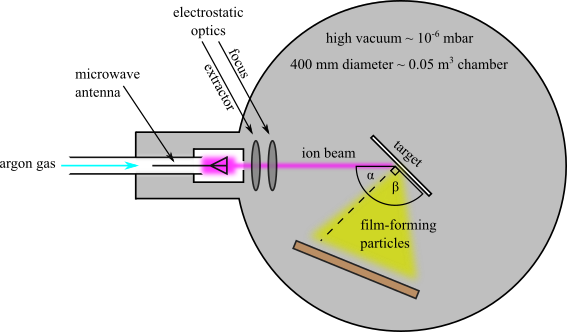}
\caption{A detailed schematic of the ECR-IBSD coating system used in this study.}
\label{schematic}
\end{figure}

The DLC coatings were deposited by a custom-built ECR-IBD system presented in Figure \ref{schematic}. The ECR ion source is manufactured by Polygon Physics \cite{Polygon}.

The source itself comprises a 1/4 wavelength resonant cavity with permanent magnet and a low power microwave antenna to produce a highly ionised argon plasma at a low pressure in the range of 3.7$\times$10$^{-5}$ to 1.2$\times$10$^{-4}$~mbar.    
Using this type of source, the plasma is generated due to ECR. In a magnetic field, electrons will move in a cyclotron motion at a certain frequency, determined by the magnetic field strength in the cavity and the mass and charge of an electron. The microwave antenna therefore imparts energy at this frequency causing the gas injected into the cavity to become ionized. In this case, a magnetic field strength of 0.0875~T is chosen and the control electronics supplies around 11.5 W of microwave power at 2.45 GHz. A beam is then extracted through a small aperture using high voltage electrostatic optics in the range of 4 to 16~keV extraction. Assuming single ionisation in the argon plasma, we attribute the nominal beam energy to the accelerating voltage applied. The focus is set to 80-90\% of this value, while the ground plate is held at 0~V. The incidence angle, $\alpha$ (normal of the target relative to the ion beam propagation direction), is set to 45$^{\circ}$. The source-target and target-sample distances were kept constant. Using this technique it is possible to produce an energetic, high purity ion beam, free from contact with an ionizing element, filament, or extraction grid as used in other sputtering techniques (such as DC Kaufman-type or RF-type sources \cite{Bundesmann2018}).
\newline
\noindent The beam is directed towards a 100 mm sputtering target at the center of the vacuum chamber where the deposition plume is directed towards the substrates. The substrates chosen in this study were 20 mm diameter JGS-1 fused silica and 316L grade 13$\times$13~mm stainless steel samples. A high purity (99.99\%), highly ordered pyrolytic graphite (HOPG) sputtering target (Scotech, Scotland) was used. The substrates are cleaned before deposition by sonicating for 10 minutes in acetone and a further 10 minutes in isopropyl alcohol. The substrates are then removed from the solvent and dried with nitrogen gun to help remove any remaining dust particulates.      
            
\subsection{Characterisation}
X-ray photoelectron spectroscopy (XPS) was used to measure the chemical composition and extract the \textit{$sp^{3}$} content. The electron spectrometer is equipped with a hemispherical analyser (Leybold EA10/100 MCD) and a non monochromatized Mg K$\alpha$ X-ray source ($h\nu$ = 1253.6~eV) was used for core level spectroscopy. The binding energy scale was calibrated using the Au 4f7/2 line of a cleaned gold sample at 84.0~eV. The fitting was performed using Unifit 2015. The fitting procedure of core level lines is described in \cite{Eren2011}.

\noindent The optical transmittance and reflectance of samples deposited on JGS were measured using a Photon RT spectrophotometer (EssentOptics LTD, Lithuania) in a range of $\lambda$ = 185 to 2500 nm. Using the Beer Lambert relation, Tauc plots were created by first plotting a graph called “index $r$”, which is $\textrm{ln}(\alpha)$ versus $\textrm{ln}(h \nu)$, the optical absorption and photon energy respectively. The slope of the linear portion of this graph gives the value of $r$, which indicates the nature of the band gap transition. The appropriate Tauc plots are produced depending on the results of index $r$. The Tauc plot is then $(\alpha h \nu)^{1/r} \; vs. \; h \nu$. The $x$-intercept of a fit to the linear region of these plots is an estimate of the optical bandgap of amorphous semiconductors \cite{Deraman2014,Tauc1974}. Fitting was performed using OriginPro 2021 to maximise the $R^2$ value. Given the assumption of negligible optical scatter in this context, the absorption of the samples is defined as $A = 1 - (T + R)$. The optical absorption is calculated from the transmission ($T$) and reflection ($R$) coefficients through the following (Beer-Lambert) relation:

            \begin{equation}
                \alpha(\lambda) = \frac{-\textrm{ln}(\frac{T(\lambda)}{1-R(\lambda)})}{d} .
            \end{equation}

\noindent Urbach energy can also be estimated by plotting $\textrm{ln} (\alpha)$ against $h \nu$ and taking $1/\textrm{slope}$ of a linear fit to this graph. The deviation of this plot from a straight line at low energies is due to Urbach behaviour - broadening of the band tail.
        
\noindent Scout \cite{Theiss} is used to analyse the measured $T$ and $R$ data and derive an optical constant model based on a combination of fitting functions related to properties of the material. In this case, the data is fitted with terms for: ($i$) a constant refractive index; ($ii$) a harmonic oscillator for the deep UV range; and ($iii$) an OJL term for the band gap of an amorphous semiconductor \cite{OLeary1997}. Using a $\chi^2$ fitting algorithm, the fit of these parameters to the $T$ and $R$ spectra can give an optical function model for the complex refractive index, $n + ik$. Fits are performed to minimise the fit deviation.

\noindent The sheet resistance is measured using a four point probe head (Ossila, UK) in combination with a DMM or electrometer. The Ossila probe has rounded, gold-plated contacts which are spring-mounted to apply a constant contact force of 60 g to the sample, minimising damage to the coating. The probe spacing is 1.27 mm in a collinear arrangement. If the resistance is in the range of around $10^{-3}$ to $10^8 \; \Omega$, a Keithley 2000 DMM was used; if the resistance is out of the range of this device (up to $10^{15} \; \Omega$) a Keithley 6517b electrometer was used, albeit in a different wiring configuration, with only the inner two of the four contacts being used. Actually, all the samples measured had sheet resistance values of between around $10^7$ to $10^9 \; \Omega$, therefore not requiring the Kelvin method to measure very low resistance, so only 2-wire measurements were taken for this sample set. Measurements were taken in triplicate at various points across the surface of the coating.
        
\section{Results and discussion}
\subsection{Structural properties of the deposited DLCs}
\begin{figure}[H]
\centering
\includegraphics[width=.7\textwidth]{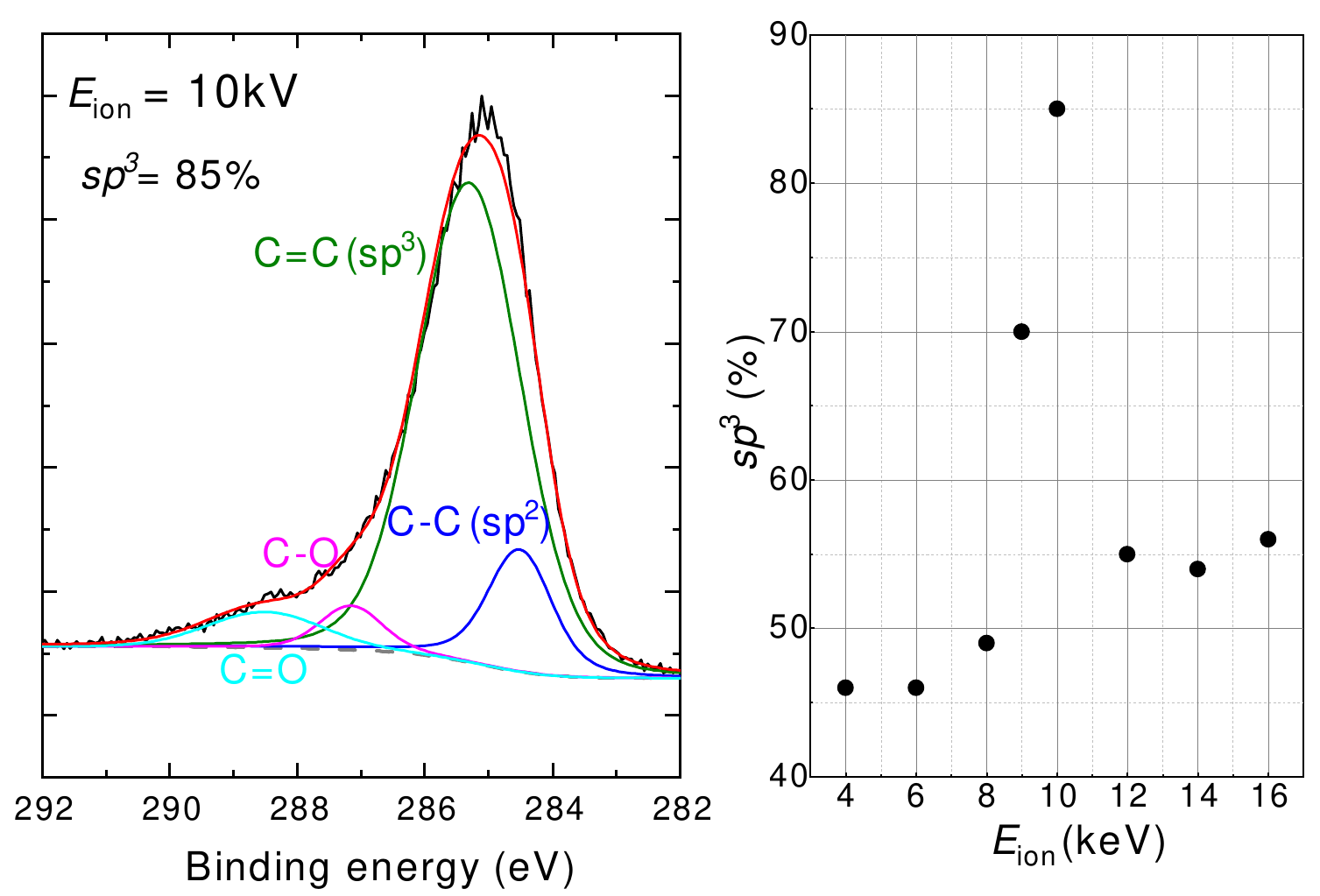}
\caption{(left) The C1s peak from XPS spectrum acquired for a sample deposited at 10~keV extraction potential. The black curve represents the raw measured data; blue, green, cyan, and magenta fitted curves are the individual chemical states; and red curves is the sum curve. (right) \textit{$sp^{3}$} content derived from the ratio of the areas of the C-C and C=C peaks for DLCs deposited at different energies}
\label{xps}
\end{figure}

Figure \ref{xps} represents the C1s core level spectra measured for the sample deposited at 10~keV extraction potential (corresponding to 4~keV ion energy for a single ionised beam). The C1s core level peak was deconvoluted into 4 singlets. The peaks at 286.8~eV and 288.5~eV are attributed to C-O and C=O \cite{Lomon2018} while the peaks at 284.6~eV and 285.5~eV are assigned to the \textit{$sp^{2}$} and \textit{$sp^{3}$} components of the DLC \cite{Li2018}. XPS measurements also revealed the existence of carbon and oxygen. The presence of the oxygen on the surface is due to air exposure of the samples after the deposition.

Although not shown here, XPS was used to measure all samples deposited at different ion energies from 4 keV to 16 keV. The \textit{$sp^{3}$} content was derived from the ratio of the areas of the C-C and C=C peaks in the XPS spectra and results are shown in Figure \ref{xps} (plot of the right). The  \textit{$sp^{3}$} ranged between 45 and 85\% where the maximum value corresponds to samples deposited at 10~keV. The energy of the film-forming particles sputtered from the target is not know unlike the energy of the Ar accelerated ions. The optimal 10~keV ion energy can be associated with the generation of film-forming particles with enough energy for densification of the forming coating for producing a high proportion of \textit{$sp^{3}$} bonds but not too high to cause thermal dissipation that results in a lower fraction of \textit{$sp^{3}$} bonds \cite{Robertson2002, Lifshitz1989}. The maximum \textit{$sp^{3}$} fraction of 85\% at 10~keV is notably high compared to \textit{$sp^{3}$} fraction in DLCs manufactured by other sputtering techniques which generally produces an \textit{$sp^{3}$} content lower than 50\% \cite{Vetter2014}. Also, it is worth mentioning here that all coatings were manufactured at room temperature which allows more flexibility in the choice substrate to use (insulating, temperature sensitive, etc…) for producing high \textit{$sp^{3}$} DLCs for different applications.

\subsection{Optical properties of the ECR-IBD deposited DLCs}
\begin{figure}[H]
            \centering
            \includegraphics[width=\textwidth]{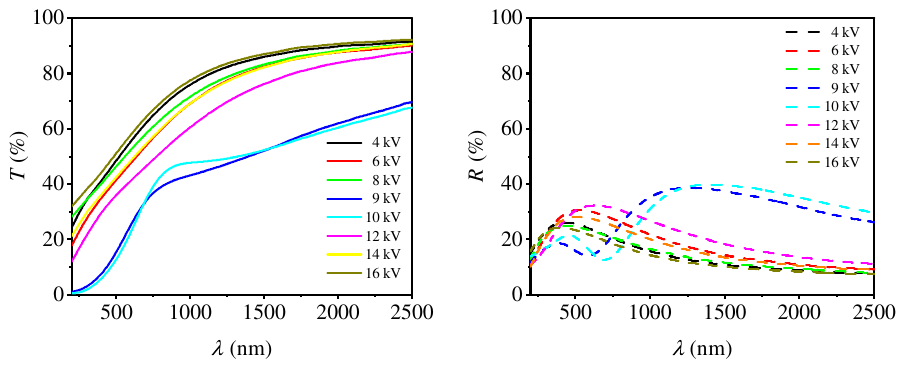}
            \caption{Transmittance (left) and reflectance (right) measured from $\lambda$ = 185  nm to 2500 nm.}
            \label{tr spectra}
            \end{figure}

\paragraph{Transmittance, reflectance, refractive index, extinction coefficient}
The $T$ and $R$ data for the DLC coatings are shown in Figure \ref{tr spectra}. The transmittance spectra are shown as solid lines on the left and reflectance as dashed lines on the right. Spectra are in the range $\lambda$ = 185 - 2500 nm (185 nm being the shortest wavelength capability of the spectrophotometer used to acquire the measurements). The overall transmittance of the coatings decreases towards the 9 and 10~keV samples (blue and cyan in the figure) and the reflectance spectra exhibit an interference pattern in these two samples that is not present in the others. The transmittance at 550~nm ranges from 52.1\% at 4~keV to 17.4\% at 10~keV and increasing towards 54.6\% at 16~keV.

\begin{figure}[H]
                \centering
                \includegraphics[width=\textwidth]{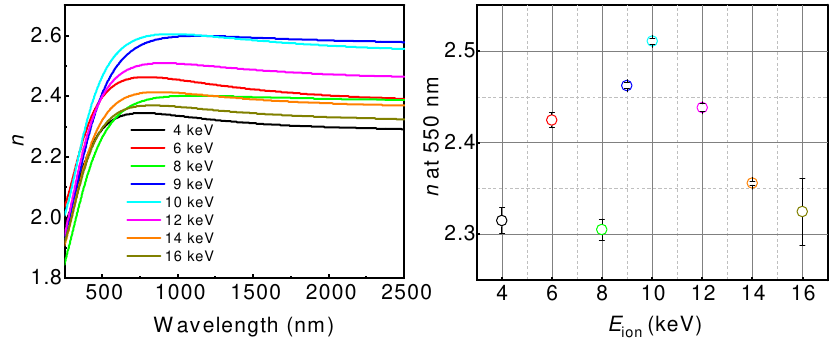}
                \caption{Refractive index spectra from $\lambda$ = 250 nm to 2500 nm (left)and at fixed $\lambda$ = 550 nm (right).}
                \label{fig:nk spectra}
            \end{figure}

Figure \ref{fig:nk spectra} presents the refractive index ($n$) and extinction coefficient ($k$) spectra obtained from Scout fittings to the $T$ and $R$ data shown earlier. The spectra are in the same wavelength range as before and presented with the same colour and solid/dashed line scheme. The refractive index (left) can be seen to be highest again at 9 and 10~keV (blue and cyan).

\begin{figure}[H]
                \centering
                \includegraphics[width=\textwidth]{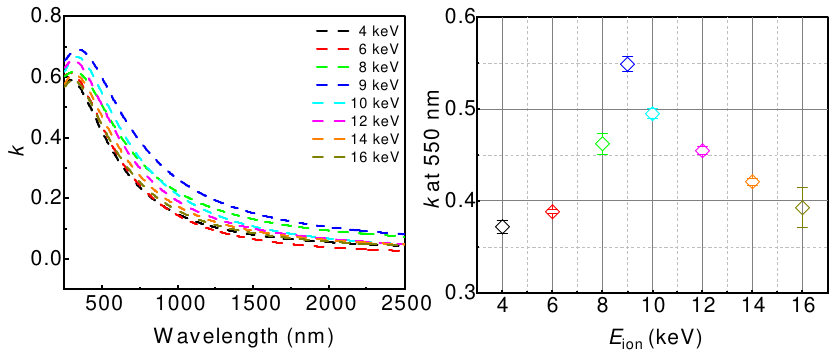}
                 \caption{Extinction coefficient spectra from $\lambda$ = 250 nm to 2500 nm (left)and at fixed $\lambda$ = 550 nm (right).}
                \label{fig:nk scatter}
            \end{figure}

Scatter plots showing the $n$ and $k$ values at 550 nm are also shown, in figure \ref{fig:nk scatter}. Both are seen to start low at low ion energy values and increase towards maxima at 9 or 10~keV and then decrease again towards higher ion energies. Refractive index (at 550 nm) are all in the range 2.30 to 2.51 and extinction coefficient range from 0.37 to 0.55. The refractive index is 2.31 at 4~keV and 2.32 at 16~keV and the maximum of 2.51 occurs at 10 keV deposition. The extinction coefficients are 0.37 at 4~keV, 0.39 at 16~keV, and the maximum 0.55 occurs at 9~keV deposition.

The transmittance and reflectance, refractive index and extinction coefficient of these samples reveal spectra and values within similar ranges to DLC coatings produced by other methods and shown in the majority of other research on DLCs referenced in this paper. These values and the trends within the data for these quantities offer some insight into the overall process when combined with the other data in this study. All results indicate a similar trend of measured properties having a maximum/minimum at the middle values of ion energy studied (9-10~keV) and decreasing/increasing (respectively) towards the outer ion energy ranges.

\paragraph{Optical band gap and Urbach energy}
                
\begin{figure}[H]
                \centering
                \includegraphics[width=\textwidth]{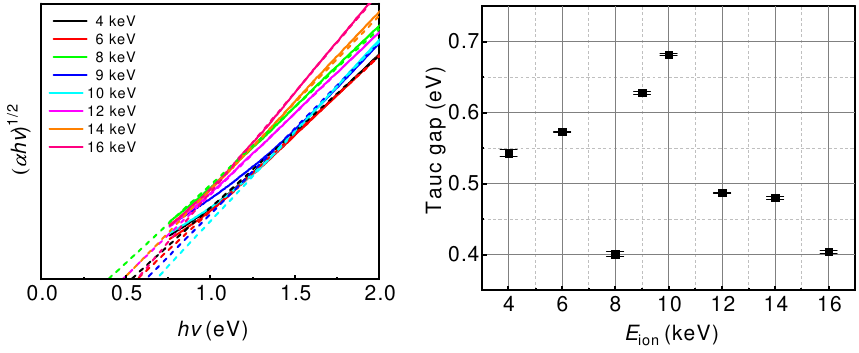}
                \caption{Tauc optical gap determined from a linear fit to the low energy region of the plot $(\alpha h \nu)^{1/2} \; vs. \; h \nu$ calculated from the Beer-Lambert relation (left) and the values for the band gap obtained by taking the $x$-intercept of these linear fits.}
                \label{tauc}
            \end{figure}

Figure \ref{tauc} shows the (Tauc) optical band gap ($E_{\textrm{gap}}$) results obtained from the method described in section \ref{experimental}. These results show a trend increasing from lower ion energies to their highest value at 10~keV and decreasing again towards higher ion energies. The lowest value for $E_{\textrm{gap}}$ is from the sample deposited at  8~keV of 0.4~eV and the highest value of 0.68~eV is at 10~keV. These fittings were performed according to the procedure in \cite{Deraman2014}, first obtaining index-$r$ and fitting to a plot where the $y$-axis ($\alpha h \nu ^{1/r}$) has the appropriate index. From these index-$r$ fits, the closest value (with highest $R^2$) across all samples was found to be $r$ = 2; therefore, we see an indirect allowed transition.

\begin{figure}[H]
                \centering
                \includegraphics[width=\textwidth]{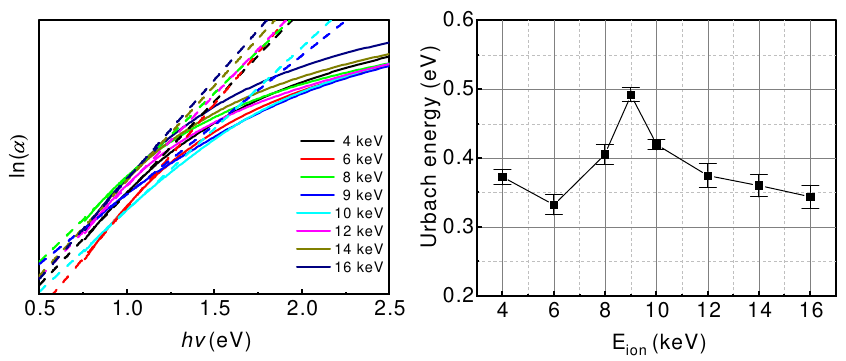}
                \caption{Urbach energy determined from a linear fit to the low energy region of the plot $\mathrm{ln}(\alpha) \; vs. \; h \nu$ calculated from the Beer-Lambert relation (left) and the values for the Urbach energy obtained by taking 1/slope of these linear fits.}
                \label{urbach}
            \end{figure}

Figure \ref{urbach} shows the Urbach energy (also obtained from the Tauc method). These are obtained from the slope of a linear fit of ln($\alpha$) vs the photon energy, also described by \cite{Deraman2014}. These results also more or less show a relationship of increasing from low values at low ion energy to a maximum at the 9~keV mark and decreasing again towards higher ion energy. All values are between 0.33 eV and 0.49 eV. At 4~keV the Urbach energy is 0.37 eV increasing to 0.49 eV at 9~keV and decreasing towards 0.34 at 16~keV. The minimum is 0.33 eV for the sample deposited at 6~keV. The Urbach energy values for these samples are comparable to the $E_{\textrm{gap}}$ values for the indirect allowed transitions obtained for the DLC coatings.

Interestingly, the band gap transitions in these DLC coatings obey an indirect allowed nature and the Urbach energies present are of similar scale to the band gap values. DLC thin films are usually reported as having somewhat higher band gap values ($\sim$ 2 eV, \cite{Robertson1997}) often being obtained using an $r$ value of 1/2 but with no analysis of index-$r$, which has been done in this study. The relatively large values for Urbach energy are not entirely unexpected as this quantity is related to highly disordered systems, amorphous materials having a higher density of states inside tails that decay into the band gap region rather than in crystalline materials which tend to have a more sudden onset of the absorption region.

\subsection{Electrical properties}
\begin{figure}[H]
            \centering
            \includegraphics[width=0.5\textwidth]{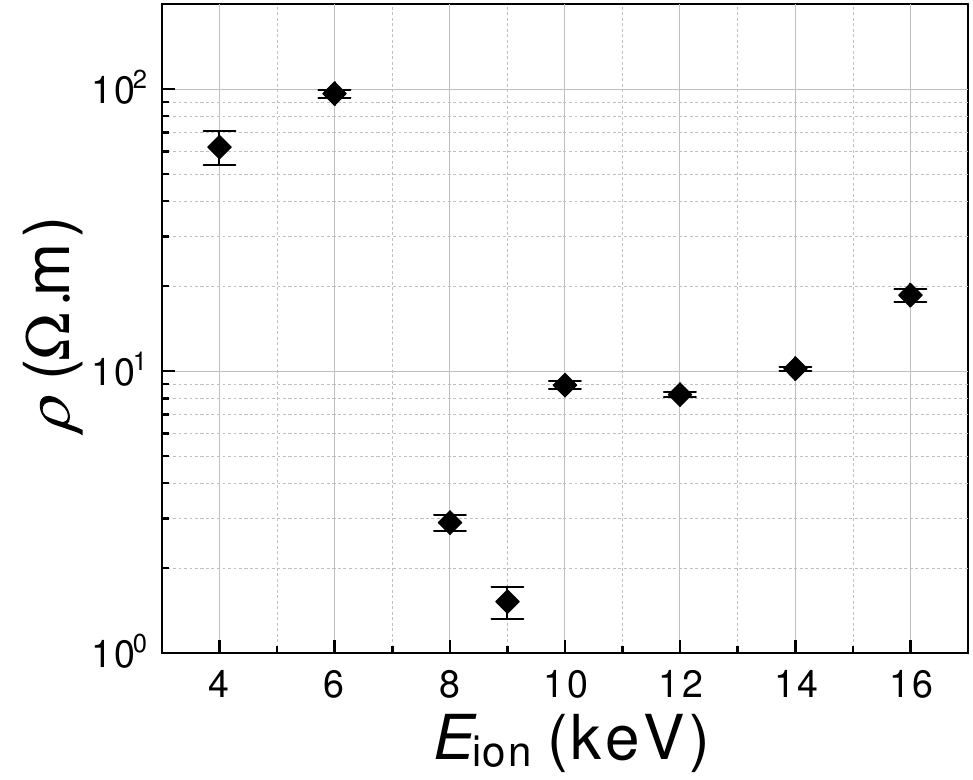}
            \caption{Resistivity of DLC coatings  produced at different energies.}
            \label{fig:resistivity}
        \end{figure}

The resistivity values are shown in figure \ref{fig:resistivity}.The $y$-axis is shown on a logarithmic scale as is often convention for resistivity data, since the values are spread over a wide range. The resistivity data are calculated from:

        \begin{equation}
            \rho = R_\textrm{s} \cdot d
        \end{equation}

Where $\rho$ is the resistivity, $R_\textrm{s}$ is the measured sheet resistance, and $d$ is the thickness of the coating, obtained from the optical fitting procedure in Scout. The trend in these data go from high values at 4 and 6~keV to a minimum at 9~keV back to high at 16~keV, with resistivity values ranging from $1.5 \pm0.2 \; \Omega$.m to $96.3 \pm3.2 \; \Omega$.m at 6~keV. 

The trend in the resistivity data is opposite to the expected trend when considering that graphite is conductive and diamond is an insulator. It stands to reason that the resistivity of mixed \textit{$sp^2$}/\textit{$sp^3$} materials would reflect this. Samples having a higher proportion of \textit{$sp^3$} bonding going towards higher resistance values and samples with higher \textit{$sp^2$} fraction moving towards conductive. The resistivity of diamond, with a band gap of 5.5 eV \cite{Angus1988}, should be expected to be very high while graphite conducts electricity. These materials, however, are both crystalline. Band gaps in crystalline materials are characterised by the sharp onset of the absorption edge translated by a very small Urbach tails, therefore very low Urbach energy. It can be seen in the results for these DLC coatings that the Urbach energies are high on the order of $10^{-1}$ eV, with the highest value of almost 0.5~eV (sample deposited at 9~keV), indicating high energetic disorder in this sample. This is also where the resistivity in the samples is the lowest. This can be understood by the small band gap and relatively large Urbach energy; where the so called Urbach tails, the exponential part of the absorption spectrum, extend far into the band gap region contributing to a higher than usual density of states in the gap.

\section{Conclusion}
The ECR-IBSD technique has been shown to produce DLC coatings (as deposited) with proportionally higher $sp^3$ content than previously reported all for sputtering techniques. The results also show that it is possible to tune this atomic structure by changing the deposition parameters of the system, namely the ion energy or accelerating potential of the beam.
By utilizing the OJL model implemented within SCOUT software, the refractive index were found to be in the range of $n$ = 2.30 - 2.51, and the optical (Tauc) bandgap energy were found to be in the range of 0.4 - 0.68~eV for the different deposition parameters that were investigated. The results indicated that the $sp^3$ content, refractive index, band gap, and Urbach energy go to their highest values with the ion energy during deposition at 9-10~keV, whereas the resistivity decreases. 
From the optical characterizations, it was found that by controlling the ion energy during deposition, the optical properties of DLC thin films, fabricated by ECR-IBSD technique could be tuned to the desired value, depending on the application. Moreover, the exploitation of high energy sputtering processes, as shown within this study, vastly increases the range of properties achievable and can therefore open up enhanced and new opportunities for robust optical and optoelectronic application of DLCs.

\paragraph{Acknowledgments}
We are grateful for financial support from STFC (ST/V005642/1, ST/W005778/1, and ST/S001832/1) and the University of Strathclyde. S.R. is supported by a Royal Society Industry Fellowship (INF$\backslash$R1$\backslash$201072). We are grateful to the National Manufacturing Institute for Scotland (NMIS) for support, and we thank our colleagues within SUPA for their interest in this work.

\paragraph{Author contributions} \textbf{C. Wiseman}: study conception and
design, data collection, analysis and interpretation of results, writing and editing. \textbf{M. Ben Yaala}: study conception and design, data collection, analysis and interpretation of results, writing and editing, supervision. \textbf{C. Gier}: interpretation of results, review and editing, \textbf{L. Marot}:data collection, review and editing, \textbf{C. McCormick}: review and editing. \textbf{C.Clark}: interpretation of results, review and editing. \textbf{S. Rowan}: funding acquisition. \textbf{S. Reid}: funding acquisition, supervision, review and editing

\paragraph{Conflicts of interest or competing interests} Not Applicable

\paragraph{Data availability} The datasets used and/or analysed during the current study are available from the corresponding author on reasonable request. All data generated or analysed during this study are included in this published article.



\bibliography{sn-bibliography}

\end{document}